# A Primer on Gibsonian Information

Bradly Alicea[1,2,3], Daniela Cialfi[1,4], Avery Lim[1,5], and Jesse Parent[1,5]

## Abstract

Across the scientific literature, information measurement in the nervous system is posed as a problem of information processing internal to the brain by constructs such as neuronal populations, sensory surprise, or cognitive models. Application of information theory in the nervous system has focused on measuring phenomena such as capacity and integration. Yet the ecological perspective suggests that information is a product of active perception and interactions with the environment. Here, we propose Gibsonian Information (GI), relevant to both the study of cognitive agents and single cell systems that exhibit cognitive behaviors. We propose a formal model of GI that characterizes how agents extract environmental information in a dynamic fashion. GI demonstrates how sensory information guides information processing within individual nervous system representations of motion and continuous multisensory integration, as well as representations that guide collective behaviors. GI is useful for understanding first-order sensory inputs in terms of agent interactions with naturalistic contexts and simple internal representations, and can be extended to cybernetic or symbolic representations. Statistical affordances, or clustered information that is spatiotemporally dependent perceptual input, facilitate extraction of GI from the environment. As a quantitative accounting of perceptual information, GI provides a means to measure a generalized indicator of nervous system input, and can be characterized by three scenarios: disjoint distributions, contingent action, and coherent movement. By applying this framework to a variety of specific contexts, including a four-channel model of multisensory embodiment, we demonstrate how GI is essential to understanding the full scope of cognitive information processing.

## Introduction

Two main features of animal nervous systems are to process information and represent aspects of the external world. As a mathematical tool, information theory is not only relevant to how information is processed in the nervous system, but also in terms of how computational agents engage in autonomous control and decision-making (Cover and Thomas, 1994; Ben-Gal and Kagan, 2021). In a complementary manner, representations of the external world can be characterized algebraically as a series of geometric projections (Richardson and Louie, 1983) which interpret the external world from multiple but invariant perspectives. Invariant features provide a scaffolding for probability distribution characterization of the environment (Frank, 2016). For our purposes, quantifying information processing in this way enables a heterogeneous view of internal representations and/or sensory inputs. Therefore, our primary interest is in building a framework to understand the role of dynamic environmental information on shaping behaviors generated by ecological interactions with the nervous system (see Figure 1). Gibsonian Information (GI), inspired by an ecological perspective (Gibson, 1979; Lobo et.al, 2018), provides us with an alternative theoretical perspective, emphasizing the role of information content in the relationship between ecology, nervous systems, and naturalistic interactions. GI provides a quantitative framework of information processing as a continuous dynamical phenomenon.

---

[1] Orthogonal Research and Education Laboratory, Champaign-Urbana, IL and Worldwide bradly.alicea@outlook.com
[2] OpenWorm Foundation, Boston, MA
[3] University of Illinois, Urbana-Champaign, Urbana, IL
[4] University of Chieti-Pescara, Pescara Italy, danielaciaifi@gmail.com
[5] Plot Twisters, https://www.plottwisters.org/

Taking an ecological perspective allows us to approach information processing from a continuous perspective rather than a cognitive one (Lobo et.al, 2018). Rather than focusing on information processing as being solely a property of the mind, ecological models require considering the agent and environment as a unified system (Corris, 2020). J.J. Gibson (1950) defines the perceptible properties of the environment as summarizable by higher-order spatiotemporal variables (see also Warren, 2021). Phenomenologically, this can take the form of biological motion and its physical correlates (Johansson, 1973; Callan et.al, 2017; Wand et.al, 2022), or optic flow during naturalistic behavior (Matthis et.al, 2022). GI assumes an embodied observer, alignment among multiple sensory modalities, and coordinated collective behavior. The calculation of GI involves the evaluation of exponential spatiotemporal distributions (which reveal statistical affordances) by a mathematical or symbolic model. Thus, a full account of GI includes both a minimal representation with feedback for generic perception and a higher-level symbolic representation that accounts for multiple sensory channels. Clusters of GI in a spatiotemporal space form statistically-defined affordances. Many times these correspond to well-categorized objects such as a door handle or rotating propeller, but sometimes can be novel clusters that correspond to spurious relationships. In its operational form $g(i)$ (and unlike Shannon information), much of the information content of GI is redundant (Zavagno, 2023). However, this redundancy is crucial for internal representations and behavioral generation. This can not only lead to the incorporation of objects that have not been observed previously, but can also lead to sensory illusions and spurious correlations (Sherrick, 1968; Stocker and Simonelli, 2006; Miyazaki et.al, 2010; Watanabe et.al, 2014; Weilhammer et.al, 2017; van der Wal et.al, 2018; Leptourgos et.al, 2020).

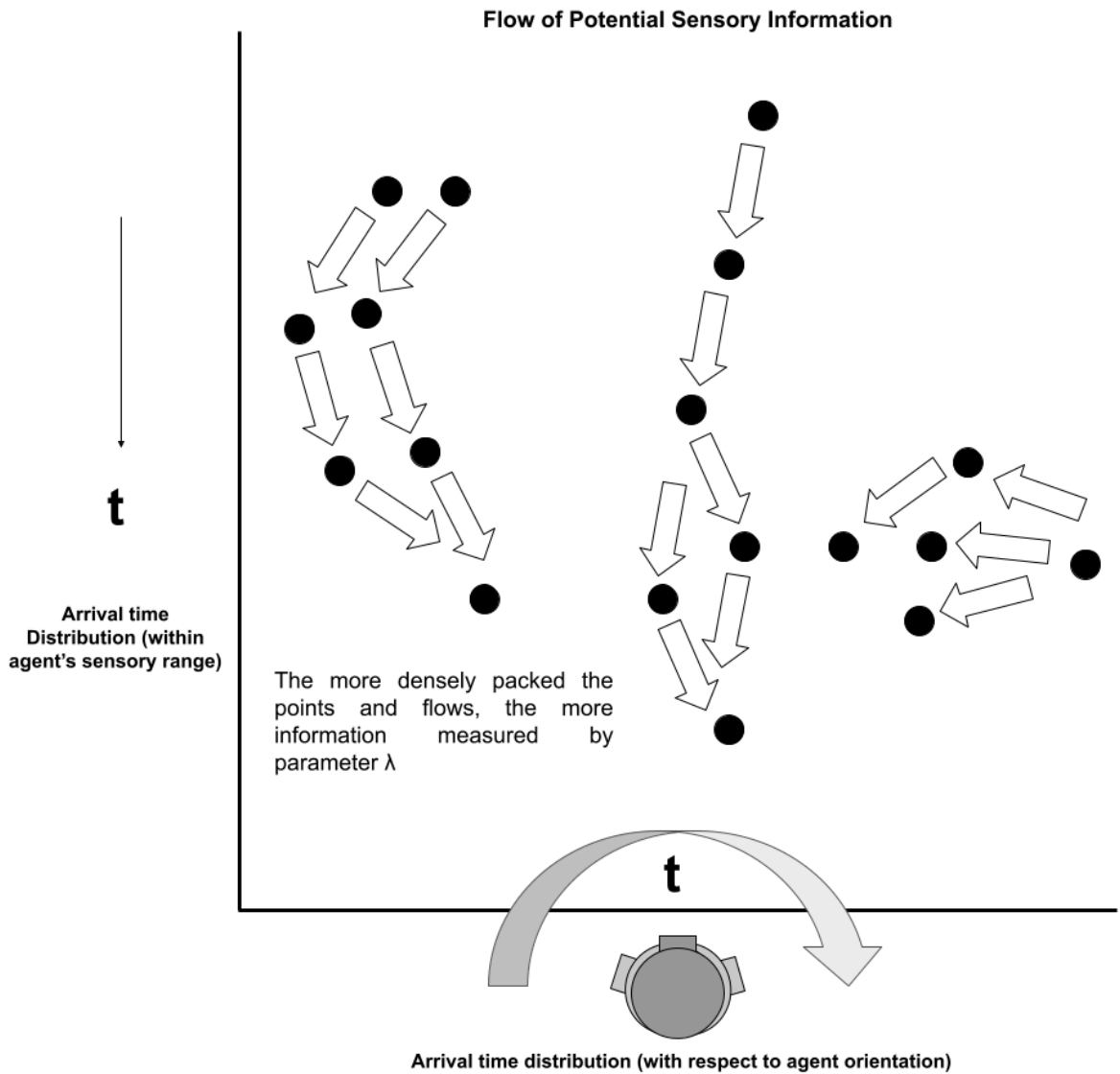

Figure 1. A computational agent that observes its environment from a stationary perspective. Discrete stimuli (black points) move towards the agent along the transparent arrows. Black points are Poisson distributed, which describes its arrival time distribution. Features accumulate as a function of λ. Each arrival adds to our cumulative signal, represented by our spatial volume for coherent movement *c*. The delay $\tau$ adds temporal noise.

GI is an inherently multisensory approach, and often relies upon synchronized multisensory integration (Bertelson and DeGelder, 2003; Elshout et.al, 2020). In the peripheral nervous system, GI can provide information about autonomous nervous system function and stress physiology (Zimmer et.al, 2022). In non-neuronal systems, various signaling mechanisms can supplant the role of neuronal sensation. In collective contexts where there are many agents, the measurement of the spatial and temporal frame of one observer can be different from that of another agent. This heterogeneity of GI across agents may predict how order emerges from spontaneously formed collectives.

GI provides a unique interpretation of information theory in addition to an alternative framework that stresses the observer-centric and affordance-based aspects of perceptual information. Whereas Shannon information relies upon the presence of variation, GI instead relies upon the presence of interpretable motion against a background. This is similar to what is captured by O-information (Scargliarni et.al, 2023), where information consists of redundant and synergistic components detected against a series of physical gradients. The relationship between motion and background are potentially differentiable in a number of ways: the mode of difference is left up to the individual observer. From the perspective of information processing in agents, this can be summarized as direct perception that is detectable in systems as diverse as seeing biological motion in humans (e.g. Johanssen, 1973) and electroreception of prey in sharks (e.g. Hopkins, 2010). Yet the non-reliance of GI on internal representations may even allow us a mechanism for grounding semantic content (Harnad, 1990) in interactions with affordances and their spatiotemporally dependent context.

**What is our focus and who is GI for?**

The GI approach can be applied to many different types of agent, from multicellular organisms with nervous systems to biological cells, and from computational agents to multi-agent collectives. GI is appropriate for information processing broadly construed, although most relevant to animal nervous systems and bio-inspired artificial systems. Given the general nature of our proposed internal model, the GI paradigm is meant to encompass multicellular animals and plants with complex perceptual systems, single cell organisms and colonies, embodied computational agents, and robots. This provides a means to understand active perception as a universal phenomenon, and to apply GI at different degrees of complexity. We use the term agent to refer to both organisms and computational agents. This does not imply any greater agency than the term organism or autonomous system that are used in the literature. Rather, our model is meant to bridge the variety of contexts where information processing is essential for understanding continuous interactions with the environment. We provide a variety of examples ranging from basic psychophysics to the perception of swarms and flocks to provide examples for a wide range of applications. Our mathematical framework also provides a path to constructing more detailed dynamical systems, agent-based, and experimentally-tractable models that can enable the measurement and analysis of GI.

**Definition of Ecological Information**

Defining information in the ecological sense (GI) is much different than a purely statistical or algorithmic one. The evaluation of GI relies on objects that are subject to apparent motion (in any sensory modality) as an agent moves through and changes perspectives on its environment. These objects can be considered affordances (Chemero, 2010), the information for which is contained in events that result in conjunction with agent interactions. These events occur in both time and space, and occur both externally and internally to the agent (Gibson, 2000). GI also provides us with possibilities of action given the apparent movement of objects relative to agent interactions (Rietveld and Kiverstein, 2014). This provides us with a generative model of information that allows for information to be extracted from time-dependent, collective behavioral, and statistically abnormal contexts.

Gibson (1954) provides a host of transformation rules for perceiving different types of motion in terms of their geometric physics, including the translation and rotation of rigidity, size transformation and deformations related to elasticity, and disjunctive movements. While the last of these (disjunctive movements) is directly related to GI, understanding the biological underpinnings of GI can be summarized through two perceptual mechanisms. Stimulus contrast reversals (overlapping bars oriented in space) and being embedded in a moving object (train moving down the tracks) provide a means to separate patterns of sensory (visual) stimuli and physical displacement. In terms of GI, this is summarized as two differential sources of $g(i)$ or a single channel of collective movement. Stimulus contrast reversals can be summarized as the following: a pair of random dots presented in a certain way can simulate motion while background cues remain uncorrelated (Sato, 1989). When a set of dots or bars are overlapping and displaced in one direction, perceived motion is detected in the other direction (Mather and Murdoch, 1999). This suggests a relationship between objects and background that is summarized in the form of disjoint distributions. Experiencing GI in a naturalistic context can be summarized through embeddedness, or the example of looking out the back of a train (Ashida and Kitaoka, 2023). From an egocentric perspective, the observer is stationary, while the background appears to move away from this viewpoint. Yet the background environment is actually the stationary component of this observer-environment relationship, marked by visual flow and the relative stability of the train on a set of tracks. Therefore, a major source of GI in this relationship is temporal change across modalities. Gibson (1979) suggests that the passage of time is a necessary component for the perception of physical change. As time passes, we witness apparent physical changes in the environment.

**Physical and Biological Basis of GI.** Additional linkages between traditional views of information and GI come in their relationship to entropy. In the case of GI, Swenson and Turvey (1991) point out that sensation provides a basis for action and movement, which in turn drives entropy production. As a feedback loop, the thermodynamics of sustained movement are both constrained and driven by entropy production. Direct perception serves to shape the internal representation as influenced by the energetic constraints of temporal updating and behavioral effects of the agent on that updated information. Shannon Information has been shown to have a correspondence to thermodynamic entropy as far back as Landauer (1961), who showed that information processing requires physical entropy to be produced. This suggests a significant role for external information in shaping the information content of agent perception. With these links to entropy, agent perception includes biological nervous systems, biological information processing in the absence of a formal nervous system, and computational agents such as embodied robots and simulations.

GI may also be important in the study of non-neuronal systems as found in slime molds (Boisseau et.al, 2016), the ciliate protozoan *Stentor* (Dexter et.al, 2019), and any number of unicellular organisms (Reina et.al, 2018). GI is particularly compatible with an understanding of such systems, as ecological approaches do not presume any specific type of internal representation or computational mechanism (Fultot et.al, 2019). These systems, characterized by both the absence of a central nervous system and the presence of a body (cell), demonstrates the essential role of embodied phenotypes for functions that are analogous to information processing and even decision-making processes. These types of systems operate on information relevant to both individuals and colonial groups. Overall, GI is broadly applicable to the sensory worlds of individuals and social groups. The influence of GI on social groups may also include the ability of agents to interpret their place in collectives such as schools, herds, flocks, and swarms (Toner and Tu, 1998; Chiovaro and Paxton, 2020).

**GI as an Ecological Approach**
While GI is largely inspired by ecological and naturalistic approaches to behavior, it is also relevant to understanding emergent collective behaviors and fluid dynamics, combining an agent-based perspective with an analysis of environmental emergence and variability. This connects to non-reductionist materialist approaches to scientific inquiry in that dynamical change is accompanied by an underlying persistence (Kugler and Turvey, 1987). This can be operationalized as persistence-change pairings, or ecological events that relate to action or preparing for an action (Warren and Shaw, 1985). As a measurement of environmental information, GI is the summary of potential persistence-change pairings in a cognitive system. Being a more universal perspective,

GI allows us to consider the interactions between environmental structure, discrete units, sensory environments, and individual agents.

**Prospective Control.** One canonical problem of ecological perception is the outfielder problem (McBeath et.al, 1995; Fink et.al, 2009). The outfielder problem is, in turn, a formulation of prospective control (Turvey, 1992). This is summarized in the following scenario: how does an outfielder catch a fly ball? This problem has traditionally been posed as one of continuous visual tracking and alignment with a goal-directed movement. This requires direct perception of an object (ball) against a background, a prediction of the object's physical behavior, and an appropriately matched output behavior by the agent. This has been characterized in the literature in the form of perception-action coupling (Stoffregen, 2000). In a broader dynamical context, this can be characterized as a set of coupled equations, evolution equations, or a multiplicative feedback mechanism.

Prospective control also allows us to derive future events and a possibility space. This is distinct from conventional prediction in that prospective control restricts possible routes to future events via the sensorium and multisensory integration. This is consistent with Turvey's notion of event-based pathways to action (Turvey, 1992). GI is also consistent with the more general case of catching a fly ball called linear optical trajectories (McBeath, 1995), which is the basis for tracking objects as they travel through the local environment. From a GI perspective, these types of trajectories convert temporal problems into a spatial framework:tracking features throughout time provides spatial displacement information, which in turn can be used to determine how much informational structure exists in continuous action.

**Bouncing Ball.** Another canonical problem is that of a bouncing ball on a target controlled by a perceptual agent. In this formulation, the ability to track movement involves the active stabilization of the action utilizing both active and passive control (Siegler et.al, 2010). Alternatively, the bouncing ball problem can be formulated as the observation of a bouncing ball simultaneous to an auditory beat (Huang et.al, 2018) or even simulated bouncing balls with a realistic motion trajectory (Gan et.al, 2015). The synchronization of multimodal sources involves frequency information that is similar to a control strategy. There are other information processing imperatives besides object tracking: maintaining a consistent viewpoint relative to the ball's motion and physical constraints of the ball's environment (e.g. specific gravity, surface resistance) is also imperative. Agent-environmental interactions can produce noise such as sensory undersampling/oversampling (Alicea, 2024) or differential masking between sensory modalities (Zook et.al, 2022), and can also factor into the ability to track and act upon the moving target, which can pose a number of information processing constraints.

**Spatial Integration.** A third canonical problem related more to multiple input modalities and spatial integration involves the visual perception of physics during spatial interactions. Spatiotemporal arrays serve as a means to provide continuous visual estimations of the environment relative to a fixed reference point. Observations of nearest-neighbor crowd dynamics (Warren and Dachner, 2018) and maze wayfinding (Ericson and Warren, 2010) predict that visual estimates add information to more traditional optimization models. In the case of navigation, the removal of constraints (maze boundaries) results in agents tending to violate the ordinal aspects of Euclidean structure. In GI, the relationship between spatial interactions and visual distance judgements are treated in a nondescript way. Yet GI also includes other modalities such as audition and touch. Taken as different perspectives that align continuous inputs (De Gelder and Bertelson, 2003), GI is also an indicator of sensory congruence. In cases where the direction of movement is not always uniform, the occurrence of gaps or inconsistencies in sensory integration become more likely.

### Gibsonian Information: mathematical definitions

**Potential Parameters for GI**

As is the case with Shannon Information, GI can be distilled into a set of quantitative values that are tangentially related to the original measurement of information content (see Appendix A and Equations 1-4). In the case of Shannon Information, the parameter *H* can be distilled into maximum information $H_{max}$, minimum information $H_{min}$, and channel capacity *sup I(X,Y)*. In Table 1, we show the analogous measures and their definitions between Shannon Information and GI.

**Communication Channels.** A full mathematical definition for GI can be found in Alicea et.al (2022). To review, GI requires a set of information channels: one external and the other internal. These are described in Equations 1 and 2 in Alicea et.al (2022). These are illustrated in Figure 2. Both of these channels produce an information function *i(t)* that describes the flow of information encountered in a naturalistic setting.

The content of our external communication channel can also be described in two ways: a Poisson arrival model (Equation 3 in Alicea et.al, 2022) and a differential measure (Equation 4 in Alicea et.al, 2022). These components are shown in context in Figure 1. We assume that this input is Poisson distributed because as a sensory stream it can be both irregular and stochastic. The Poisson model also introduces a temporal component with delays (τ), and describes the rate and tempo of the arrival of sensory stimuli from the environment. When a positive or negative delay (-τ, τ) is included in the distribution of environmental information, the sensory information is transformed into an expedient representation. As the observer moves throughout its environment, or conversely when the environment moves relative to a stationary agentive reference frame, rate and tempo of environmental information becomes key to sustaining perceptual coherence. Mathematically, this provides us with a time-series function (parameter $\tau$').

Table 1. Differences between Gibsonian and Shannon Information, and the method by which these differences might be measured.

| Gibsonian | | Shannon | |
|---|---|---|---|
| Flow, movement, contrast | Maximum coherent movement $(MV_{max})$ | Surprise, heterogeneity, diversity | Maximum entropy $H_{max}$ |
| Disjoint distribution (difference between moving pixels and stationary pixels over time) | Minimum coherent movement $(MV_{min})$ | Joint distribution (mutual information between pixels of different states) | Minimum entropy $H_{min}$ |
| Temporally dependent (time-series) | Flow Rate Capacity $sup\ P(\frac{e^{-\lambda}\lambda^k}{k!})$ | IID (binomial probability distribution) | Channel Capacity $sup\ I(X,Y)$ |

Since GI is dependent on the sensory organ, which does not need to be visual, Gibsonian Information is often weighted according to a center-surround model of signal transduction. For example, signals encountered at the edge of a sensory field, or in an oblique manner, are down weighted as compared to signals encountered head-on in the middle of the sensory field. This can be represented in the internal communication channel as fuzzy or uncertain information. The internal communication channel (Figure 2) holds a representation of GI as an encoded Poisson distribution with temporal transformation. This representation can then be decoded as a means to drive behavior or other autonomous processes within the agent.

Alternatively, the differential model enriches our temporal perspective with a spatial component. The static features and dynamic motion of the environment, in particular affordances, are located in a spatial context. To account for aspects of an environment that moves relative to the location (not the reference frame) of the observer, we utilize a continuous function that measures the contrast between an environment's features and their background. This provides a means to identify potential affordances, and when combined with temporal information provide distance, depth, and continuity cues. Mathematically, this can be done by finding the difference between two values of parameter $d_i$.

**Formal Parameters**

**Coherent Movement.** A typical ecological analysis of attention and crowd perception does not fully capture its collective behavioral aspects (Wilcoxon and Warren, 2022). Therefore, our definition of coherent movement is defined as an observation of all adjacent agents in a sensory field. Coherent movement is the transformation of sensory channel signals into the dimensionality of space. Given a local and current frame of reference with a given resolution and depth, coherent movement is summarized by the perceptual parameter *c*

$$c = \int_0^m s(x) \bullet \int_0^n t(x) \times \int_0^o p(x) \qquad \text{[1]}$$

where *s(x)* is *m* elements of $g(i)$ in 2-D space, *t(x)* is *n* elements of $g(i)$ over time, and *p(x)* is *o* elements of $g(i)$ in terms of depth (e.g. distance from the agent). *s(x)*, *t(x)*, and *p(x)* are collectively derived from raw environmental information. A scalar triple product is used to represent the internal integration of spatiotemporally distributions of environmental information, and thus the superadditive aspects of multimodal integration (Holmes and Spence, 2005). In turn, the resulting scalar *c* is essential for describing moments of coherent movement. This three dimensional volume (*s,t,p*) represents the attentional mechanism of the agent with respect to both collective phenomena (flocks, swarms, or topological phase transitions in matter) and gestalt perception. Coherent movement allows an agent the ability to separate emergent sensory phenomena from single objects: collective movement like biological motion and atomic movement like particle following.

**Internal-related Movement.** We define internal model-related movement (*MV*) as a combination of environmental information and force production of the agent. *MV* is the difference between the sensory milieu as captured by *c* or $g(i)$ and movement production with noise. Movement differs from collective movement in the following way: Collective movement is translation of environmental movement to information, while singular movement (*MV*) is the generation of autonomous movement from projections of environmental information. Mathematically, MV is defined as

$$\text{MV} = (\prod \vec{v}_v\theta,\ \vec{v}_a\theta,\ \vec{v}_t\theta \ast \sum \vec{v}_v\theta,\ \vec{v}_a\theta,\ \vec{v}_t\theta) - (PO_{vo} - \sigma_{po}) \qquad \textbf{[2]}$$

where environmental information consists of multiplicative and additive sensory components ($v_n$) with a positive or negative directional component. Force production is distinguished between volitional force production ($PO_{vo}$) and noisy force production ($\sigma_{po}$) at the prior time step. The theta term ($\theta$) for each sensory component represents potential curvature, which for a linear interaction is 0. In the multiplicative case, the multiplied terms should have an observed or estimated value, while the added terms are all constant at 1.0. In the additive case, the multiplied terms are all constant at 1.0, while the added terms should have an observed or estimated value.

For our purposes, we can take $MV_{min}$ and $MV_{max}$ as analogues to Shannon information theory (see Table 1). In GI, $MV$ is another way to represent the external and internal information channels as shown in Figure 2.

**Flow Rate Capacity.** The flow rate capacity ($f_{max}$) is defined as the upper bound of arriving sensory information, which is equivalent to the communication channel capacity. The size of the communication channel is the supremum of $g(i)$, and is defined mathematically as

$$f_{max} \equiv sup\ P(\frac{e^{-\lambda}\lambda^k}{k!}) \qquad \textbf{[3]}$$

where we use the Poisson distribution to represent $g(i)$. The capacity $f_{max}$ simply assumes the maximum amount of information experienced is the current flow rate.

**Decay Rates.** The decay rate is defined by $d$, which represents the loss of information over time and/or space. In a purely temporal instantiation and $f_{max} = 0$, $d$ will reduce the amount of available information in linear fashion. In a purely spatial implementation, $g(i)$ will decay as sensory stimuli extend to the edge of a sensory field. The output of coherent movement ($s,t,p$) to the sensorium can be augmented with the subtractive component $d$. In a simple 4-D space (three spatial dimensions and one of time) decay occurs linearly to the edges of the egocentric space, and decreases as new GI becomes more sparse. In a 5-D space (four dimensions of space and one of time), d of the fourth spatial dimension approaches 1, unless the agent possesses the ability to sense that dimension. Decay of $g(i)$ is also true of 3-D spaces consisting of two spatial dimensions and one of time. In this case, the source of decay is due to the sparsity of environmental information; however, an agent specialized for sensing a 2-D environment would not suffer this loss. See the Environment as Continuous Gradient for a more complete description.

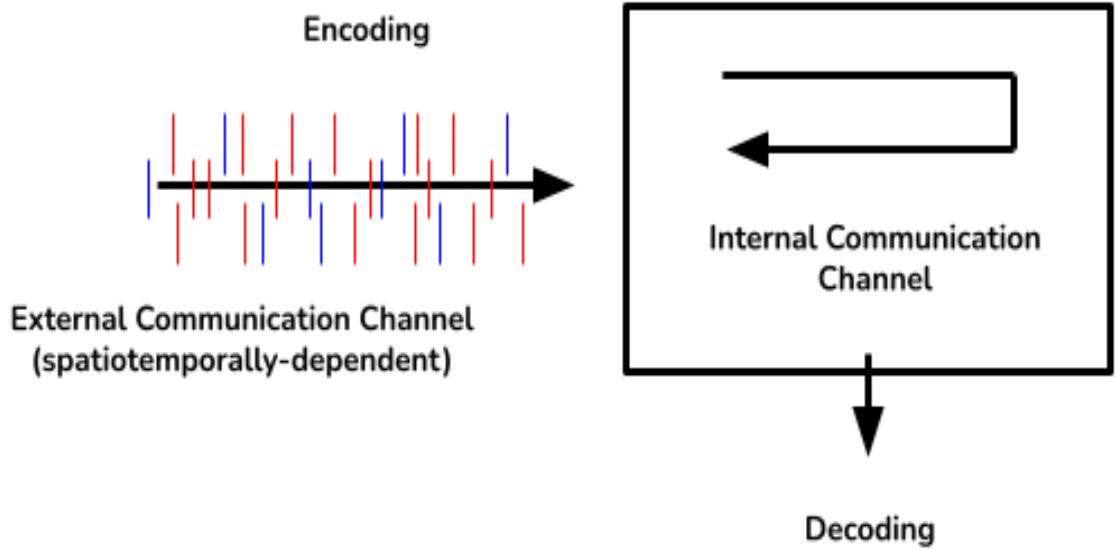

Figure 2. Perception and processing of GI in an agent. Shown are encoding and decoding steps, represented by the external environment and internal communication channel, respectively.

**Sensorium.** The sensorium (S) is defined in Figure 3, and is a combination of coherent movement (*c*, or the multiplicative combination of *s(x)*, *t(x)*, and *p(x)* scene vectors), movement output (*MV*), and information decay (*d*). The mathematical relation for the discrete and continuous cases are

$$S = (\frac{c * d}{MV}) - \tau$$

$$f(S) = \int_0^t \frac{c(d)}{MV} - \tau$$

[4]

Where $\frac{c * d}{MV}$ (or $\frac{c * d}{MV}$) is the ratio of coherent movement perceived (with decay) as a fraction of movement output generated. Information gain occurs when *S* is greater than 1.0. By contrast, information aliasing (Alicea, 2024) occurs when *S* is much less than 1.0.

**Internal Model and Representation**

Overall, our equations (1-4) suggest that an internal mechanism of some type is necessary for information processing in the agent, and results in different types of action. Direct perception itself contributes to internal information structures in a number of ways. Soatto (2009) argues that visual information is not the content of raw images, but visual structure without features such as viewpoint or illumination. GI takes viewpoint perspectives and other features of the sensory world into account, albeit as individual aspects of a complex spatiotemporal signal. The fluctuation of stimulation (and its second order correspondence with stimulation) at sensory receptors may yield all of the information anyone needs about the environment (Gibson and Gibson, 1955).

Fluctuations are demonstrated in our singular movement measure (*MV*, Equ. 2), and play an important role in the internal processing of information. Moreover, fluctuations in sensory acquisition are captured in the statistical descriptions of $g(i)$: exponential distributions rooted in both space and time that enable features to be characterized spatiotemporally. However, this is merely a necessary condition of information based on action-perception. The sensory array also provides higher-order patterns characterized by collective motion (*c*, Equ. 1) in the environment (Witt and Riley, 2014). This sensory array is linked to both internal processing (*S*, Equ. 4) and ultimately action (*MV*).

**GI as Feature-detection.** GI can be used to inform the structure of a basic feature detector. The architecture for this is shown in Figure 3. Inadvertently, this describes the relationships amongst Equations 1, 2, and 4. Equation 3 (flow rate) is simply the capacity of the main information channel (arrows from arrivals per unit time to the sensorium).

**Environment as Continuous Gradient**

This notation is for the three modality case: the direction of our sensory modality is represented by relative velocities $\vec{v}_v$, $\vec{v}_a$, $\vec{v}_t$, where the environmental source of all modalities ($\nabla v$, $\nabla a$, $\nabla t$) are moving directly against the agent (180° opposite the direction of the agent's movement). Angular displacement describes the misalignment of this gradient with respect to the observer's perspective. This is measured using a phase angle parameter ($\theta$). This allows us to describe a contour as $\nabla t(\theta)$, describing all points with set of common properties. Thus, each contour line can be described as a manifold, but for our purposes this is not necessary. Figure 4 shows this flow of stimuli relative to the agent's perspective.

In the discussion of decay rates, an agent's capacity for processing information in different spatial dimensions plays a role in accessing the full extent of environmental information. Visual range for binocular vision is described by the visual envelope (Nixon and Charles, 2017), which is a 3-D spatial volume in front of the agent. Information outside of the envelope is either not accessible, or via bodily rotation. In the latter case, there is a time delay so that the temporal transformation $\tau$ is a temporal delay. This occurs in other sensory modalities as well: weakly electric fish use an electrosensory volume (Snyder et.al, 2007) to match their sensorimotor movements to sources in the local environment, while moths engage in plume-following (Vickers et.al, 2001), which involves intricate interactions with an turbulent olfactory plume. Olfactory plumes provide higher-dimensional and even fractal spatial information.

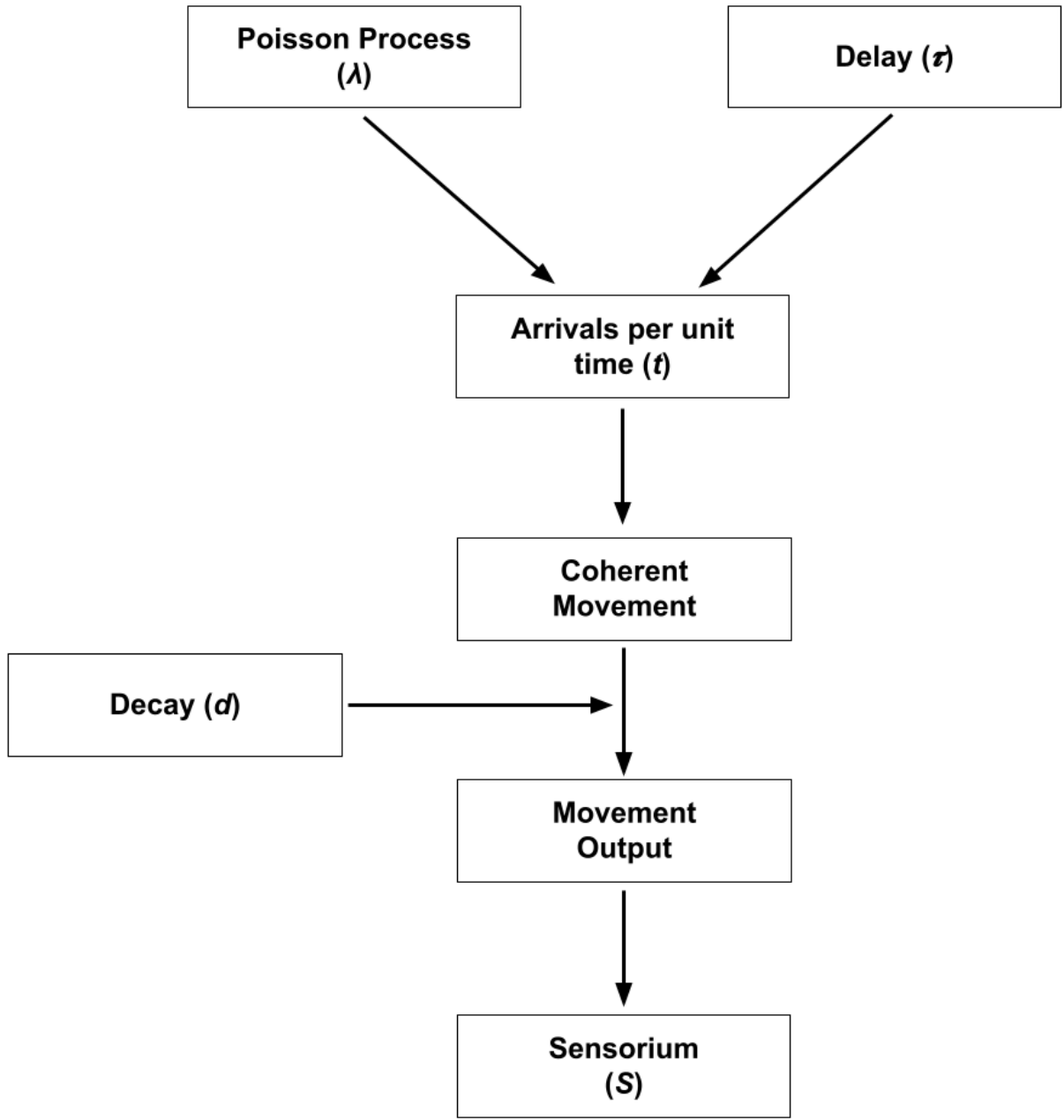

Figure 3. Feature detection in the context of GI, which results in a low-level representation called the sensorium ($S$). $S$ exists at different states in time in the form $S_t$.

Two scenarios that lead to a sparse *S* are: higher rate of decay per unit t, and a smaller λ value in conjunction with a greater lag ($\tau$).

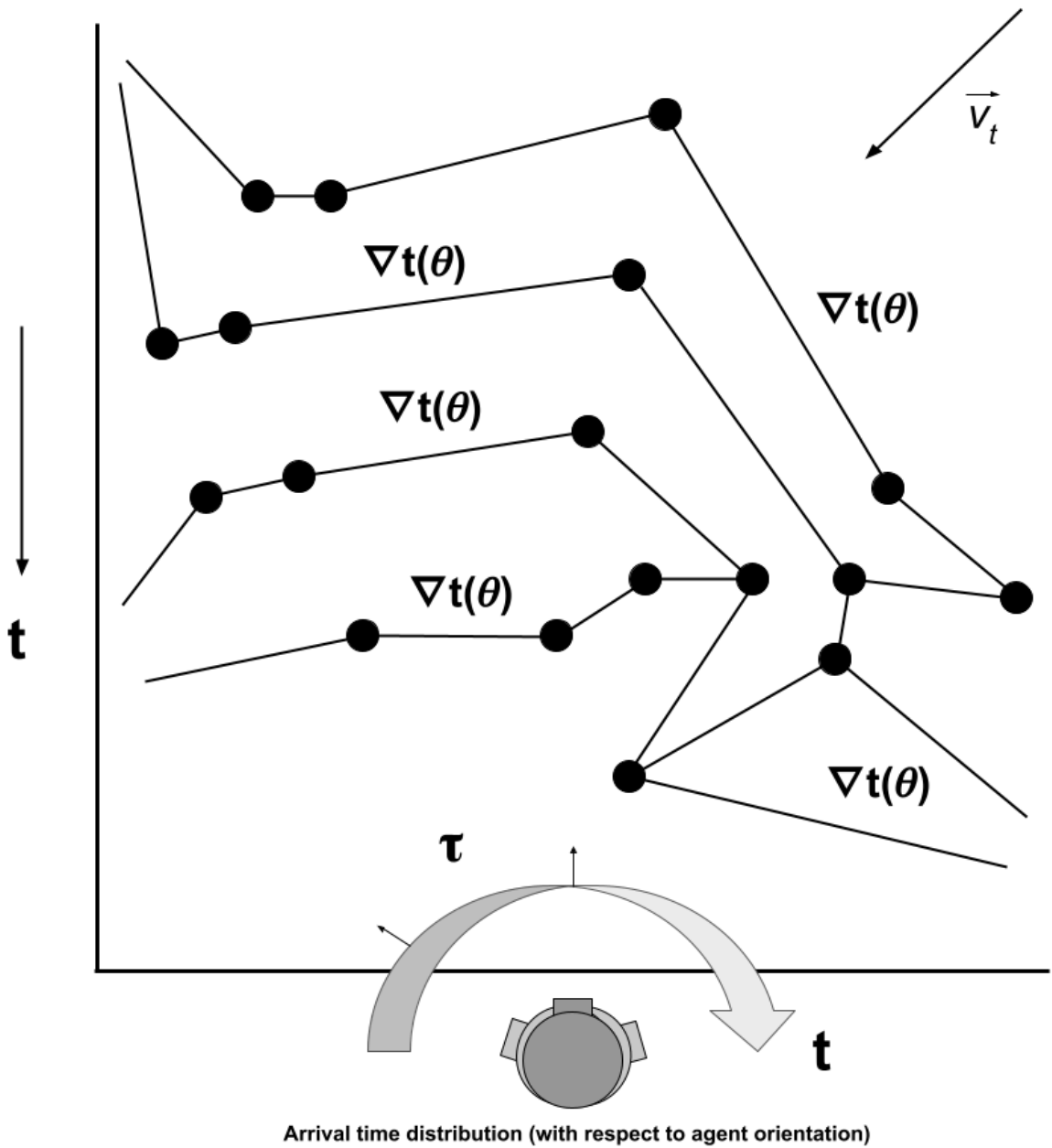

Figure 4. Contour map of the environment for $\nabla t(\theta)$. In passive sensation (e.g. Mammalian vision), environmental stimuli are perceived as a series of contours representing a sensory gradient (distance, similarity, or temperature). In active sensation (e.g. weakly electric fish electric fields), the contours are generated by the agent. Each contour is a degree of change that is computationally differentiable. The agent (agent at bottom center of the diagram) surveys their landscape using a protruding sensory organ. Surveying the environment requires attentional shifts along

the arrow in front of the agent, which takes time (*t*). Encountering different points on each contour line introduces lag (**τ**) in perceiving one part of the environment versus another.

As the capacity of sensory modalities can be represented as a volume in naturalistic contexts, multisensory integration is reliant on overlapping spatial volumes that elongate, desynchronize, and decay during free behavior. This elaborates on our concept of the sensorium (Equ. 4). The direction of interaction velocity, however, builds from our Equ. 2, where internal model-related movement engages with the edge of sensory information closest to the sensory array.

## Connections to Dynamical Systems

### Perception-action Oscillations

We can turn to a dynamical systems approach to see how sensation and action are coupled during perception. Raja (2018) provides a mathematical model suitable for discussing sensation and action as a set of coupled dynamical equations. This of course assumes an isomorphic relationship between sensation and action, where the greater congruence between the two, the less interference in this process. In our formulation of sensation and action, the example of motion can be used as an example of how these two phenomena are separable. This can lead to inversions, or anti-phase periods, between coherent movement (perception) and internal model-related movement (action). In such cases, coherent movement is the inverse of internal model-related movement, but regulates the action of an agent in more complicated ways. Interference is not a matter of noise in the loop, but a combination of flow rate capacity ($f_{max}$), decay (*d*), and delay ($\tau$).

The manner in which $\tau$ is conceived of and applied throughout the literature also differs from what is proposed here. In Raja (2018), $\tau$ is treated as a conduit to resonance, which sustains the process of cognition and perception. In general, $\tau$ is considered as a control parameter in a dynamical representation of perception-action control (Lee, 2009). Speeding up and slowing down movement relative to obstacles results in adaptive control (Fajen and Devaney, 2006). In GI, $\tau$ is strictly a delay parameter that is used in a manner similar to its use in delay differential equations, having both positive and negative components.

### Tau parameter in Dynamical Perceptual Systems

In GI, we utilize the $\tau$ parameter as temporal lag or some other form of transformation (see Figure 4). This has its origins in delay differentiation equations (DDEs), where $\tau$ is used as a lag parameter. In the Ecological Psychology context, $\tau$ is defined as time to contact (Lee, 2009; Raja, 2018). Time of contact is derived from optic flow, and does not explicitly require neither size nor velocity information. One definition involves the relative visual angle per unit time (Raja, 2018). In this definition, any estimate of visual angle is proportional to the total size of a visual field. An alternate definition given by Gibson is based on object size: the rate of expansion of the image of any point or object is inversely proportional to the distance of that point or object from the observer (Hecht and Savelsbergh, 2004). In both cases, this provides us with a volume that is consistent with our sensorium ($S$) metric. But time of contact also refers to the tracking of objects to or from the end effector (e.g. hand or foot). In our model of GI, this suggests interactions between parameters $S$, $MV$, and our conception of $\tau$.

Some of these interactions provide a natural weighting scheme that emerges from environmental features. Being a much simpler internal mechanism than the working of a complex internal model (Kawato, 1999), magnitude-sensitivity (Pirrone et.al, 2022) is an important aspect of ecological approaches. Instead of modeling inverse kinematics with a computational model of a specific brain structure (e.g. cerebellum), magnitude-sensitivity can provide simple operations to the agent critical to interpreting the incoming sensory stream. GI is premised on a similar principle: utilize the structure of the environment captured by the embodied sensory receptors of an agent to enable dynamic prediction and anticipatory behaviors. In the ecological context, spatial structure is combined with embodiment cues (e.g. body size, limb length) to self-correct movement behavior (Sridhar et.al, 2020). Our conception of $\tau$ is consistent with the latter definition in the literature, while providing a more complex spatiotemporal constant that can allow an agent to weight its interaction with the environment or other agents.

**Embodied Observers and Affordances**

One critical component of the GI approach is viewing agents as embodied observers. Observers possess a body that embeds its information processing system, and interactions with the environment that provide an explicit point of view of the world. The definition of viewpoints involves the relative congruence of sensory modalities, which further provides constraints specific to the observer's location and prior experience. Congruence of sensory modalities are determined by how an observer leverages both isolated and correlative affordances (Chong and Proctor, 2020). In addition, broader ecological and enactive approaches assume information to be both content-free and utilized in support of affordances (Hutto and Myin, 2013; van Dijk et.al, 2015). While we explicitly ignore the role of attention, it nevertheless provides a generative mechanism for encoding information from the environment into the internal communication channel.

Points of view are also critical to interpreting isolated affordances, which consist of objects such as geometric primitives, functional guideposts, or spatial wayfinding signals. In the context of GI, isolated affordances can be viewed as anomalies with respect to a random background, and serve to bootstrap the exploitation of GI by an observer. Heft (1996) provides an explicitly spatial perspective on GI. The action (movement in space) of an agent generates optical flow, which provides a continuous reciprocal interaction between the internal and external communication channels. Optical flow can be extended to flow in other sensory modalities as well: it is these multimodal, continuous aspects of the information that make GI distinct from other types of information measures.

Another important feature of GI is the utilization of what we call correlative affordances to integrate sensory inputs and make simple inferences. In the context of GI, affordances are building blocks for environmental structure. Phenomena that co-occur or are repeatedly experienced in the environment are considered correlative affordances. One example of this co-occurrence is visual confirmation of a bouncing ball coupled with proprioceptive confirmation of the ball's gravitational force. While the minimal representation required to register a correlative affordance are two sensory channels and a small embodied neural network, correlative affordances can be a gateway to associative learning. Multisensory information can also be used to facilitate the context for learning. Concurrent signals captured in different modalities, such as visual and proprioceptive features of the same object or environment, serve to provide a means to assess relative distances and sizes (Warren, 2019).

Our view of affordances is also consistent with the philosophical view that relationships between the environment and agents only make sense in context of the whole informational system (Fuchs, 2018). GI is flexible in that the observer can either watch things move around or be part of a collective that moves in synchrony. GI is embedded in space, what is relevant (affordance). Stationary aspects of the observer and environment serve as an anchor point, and govern the relationship among physical, embodied space, and movement in physical space. GI measures a change in proximity (or set of spatial relations) in relation to and between affordances.

**GI and Ecological/Embodied Resonance**

GI can also be reconciled with the notion of ecological resonance. From a historical perspective, the idea of resonance has been a poorly-conceived metaphor, serving largely as a metaphor of the nervous system "vibrating" or being "in tune" with the environment (Ryan and Gallagher, 2020). This might also be considered the radical position, and as such is not compatible with any notion of information. Gibson (1966) originally defined resonance as a system that exhibits resonance between its inputs and outputs. Yet unlike a string instrument (Shepard, 1984), the nervous system requires a distinction between signal and noise, as well as projective mappings between the central nervous system and an agent's peripheral morphology (Striedter, 2004). Other accounts target the role of brain rhythms (Hutcheon and Yarom, 2000) and connectionist networks (Grossberg, 2013) as the site of resonance. While this is more in line with the demands of neuronal mediation, this does not fully address the need to extract the informative aspects of direct perception from the environment. Constructing an explicit account of ecological information is important for enabling prediction and anticipation, not simply a time-dependent, direct mapping from stimulus to response.

More mathematically rigorous models bring us closer to a sufficient account of resonance, but do not distinguish between useful information, redundancies, and other noisy features of the nervous system-environment interaction. These are quite significant, not only in terms of stochastic resonance (good noise – see Moss et.al, 2004), but also cumulative at multiple scales. Raja (2018) utilizes the tau parameter to demonstrate the coupling of dynamical systems that can define resonance. In this case, the environment and the nervous system are captured as dynamical systems that influence each other. In a similar manner, Kelso and Engstrom (2008) argue for a coordination dynamics, which utilizes a phi parameter to describe the phase relations and potential resonance between two coupled systems. Both of these approaches describe sensorimotor systems quite well, but it is of note that sensorimotor systems explicitly involve movement and rhythm. Thus, measures of resonance are dependent on the heavily embodied (e.g. kinematic, see Shepard, 1984) closed-loop feedback so essential to sensorimotor coordination. While GI can be adapted to such systems, we have not formally proposed a method for resonance.

**GI in Naturalistic Contexts**

**GI in an R, I, E Internal Model**

One way to understand processing in an embodied agent is to think of the internal model as a three compartment model. The R, I, E model represents receptors (e.g. sensory cells), interneurons (neurons), and action effectors (e.g. motor outputs). Taken as a directed network, R, I, E constitutes the simplest embodied connectome: a sensorimotor loop and an integrator between sensor and effector. While feedback loops are secondary, the idea is to have three compartments linked by feedforward components based transformation of function. Figure 5 shows a simple three-compartment embodied GI processor based on a Braitenberg Vehicle (Braitenberg, 1984) architecture.

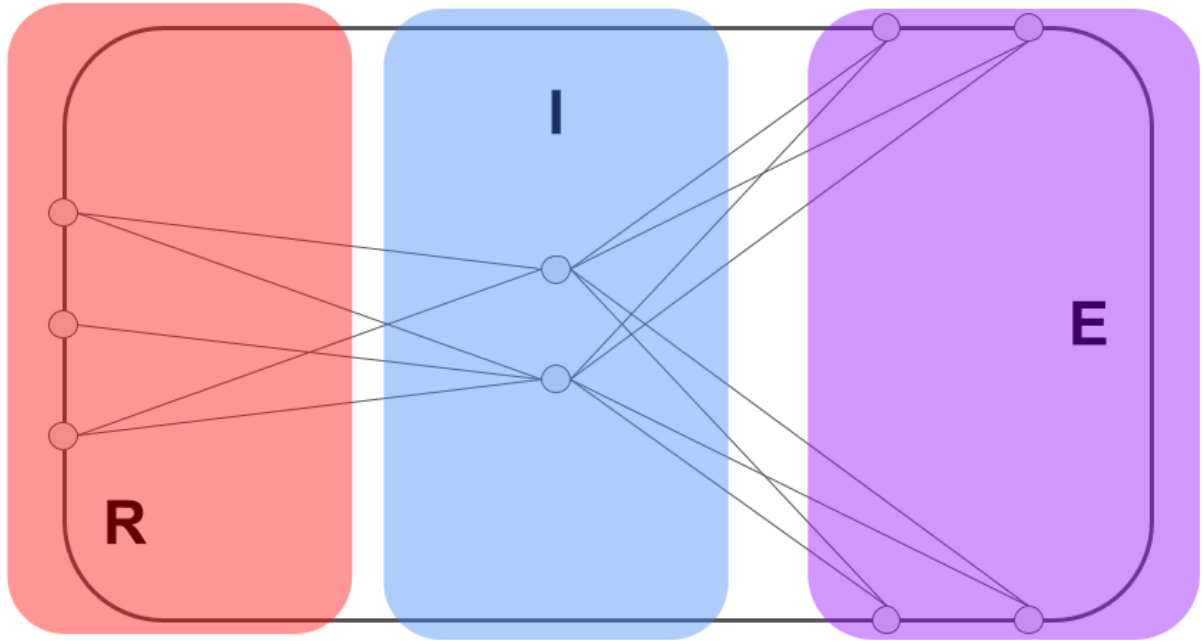

Figure 5. An example of a R, I, E three-compartment model of an embodied connectome (based on the structure of a Braitenberg Vehicle). Weighted pairwise connections between R (three nodes) and I (two nodes) and I and E (four nodes) represent the internal model structure of our GI process shown in Figure 2.

The weighted pairwise connections are defined as $GI_{i,j}$ where $i$ and $j$ represent R and I or I and E, respectively. Nodes of compartment I serve to integrate GI from different input sources. Inputs of compartment R provide spatiotemporal resolution that get integrated at each node I in the form of collective motion (parameter $c$). Integrated GI from compartment I are transformed into movement (parameter $MV$) as a feedforward signal to nodes in compartment E. The R, I, E model allows us to view direct perception from a quasi-anatomical standpoint: an internal architecture allows for behaviors to be produced from the direct perception of environmental cues such as inertial forces, light, or odorants. The existence of GI in embodied sensory channels can also be demonstrated using the developmental Braitenberg Vehicle (dBV). dBVs (Dvoretskii et.al, 2020) allow us to consider the interaction between a changing internal architecture and direct perception.

There are also relational statements that can be derived from the three-compartment model. In cases R > I and R > E, so-called rich sensory information is present in the agent. All internal processes are dominated by the sensory gradients and distributions summarized by $g(i)$. For I > R and I > E, the internal model has a high degree of integration capacity. In such cases, discrete sources of $g(i)$ will tend to be smoothed (averaged) or smeared (convoluted) over multiple inputs, but will provide highly sophisticated representations. Finally, E > I and E > R provide finer control over agent movement and subsequent spatiotemporal sampling. Examples include gradient exploration and distinguishing between signal source locations.

**GI as a Operational Phenomenon**

To better understand the GI concept as an operational phenomenon, it is necessary to shed light on three important properties that lead us to the three quantitative principles of GI. These principles are characterized in terms of the disjoint distribution of sensation, contingent action and coherent movement (Figure 6). In addition, GI can also be observed by agents in terms of bars and gratings, as well as being captured by several parameters that can be contrasted with similar parameters defined for Shannon Information.

**Disjoint distribution.** The disjoint distribution of sensation is equivalent to disjoint information. While Mutual Shannon information (Ross, 2014) is defined as correlated information content between two categories, disjoint distribution is defined as the difference between two highly similar categories. Disjoint information allows us to distinguish two categories: one that is static and one that is dynamic. It is precisely this difference that leads to the measurement of GI: The greater the apparent motion in the dynamic series of images, the greater the GI. We can use white (or Gaussian) noise to demonstrate the role of disjoint distributions in GI. White noise serves as a means to activate brain mechanisms that allow us to interpret biological motion (Callan et.al, 2017). Our example in Figure 5A shows a single frame of white noise static versus an animation of white noise static seen as a transition across multiple frames. As the difference between the two images, disjoint distributions reveal features such as the color state of specific pixel locations and motion cues. Disjoint distributions can be evaluated from both an egocentric and third-person perspective, and is best applied when a single observer is embedded in an environment separately from the influence of other observers.

**Contingent Action.** The second principle of GI can be summarized as contingent action. When making comparisons between static and dynamic categories, dynamic categories are fundamentally different because they are subject to temporal contingency (Gallistel et.al, 2013): the sequence of images that make up the dynamic comparison in our disjoint distribution example exhibit a temporal causal dependence that can be modeled using a contingency tree. This dynamic comparison can be characterized as contingent action, and is shown in Figure 6B. The GI inherent in contingent action involves a comparison of two image sequences (animation), one with missing images in the sequence, and the other with the full sequence of images intact.

This comparison should reveal that sequences with images removed from the periods of great configurational or contextual variation should yield relatively low amounts of GI.

In the example in Figure 6B, it might be enough to simply know that the Zorb ball is chasing the runners and what runners the ball rolls over. Conversely, images that contain great redundancy (conveying no transitional or configurational variation) can be removed without impacting the amount of GI. The sensitivity to configuration is comparable to Shannon Information, but Shannon Information does not explicitly deal with ordered motion as historical contingency. Long-term sensory augmentation can also result in contingencies which occur at a longer timescale (Konig et.al, 2016), but affect direct perception nonetheless.

**Coherent Movement.** Our final principle is called coherent movement. This can be demonstrated using a model of either being a participant in collective behavior, or being an outside observer of collective behavior. According to Kameda et.al (2022), individual heterogeneity and different styles of collectivized decision-making play a role in how the dynamics of a collective are acted upon by the observer. Yet we also notice that while the disjoint distribution measure makes a comparison between static and dynamic scenes, it says nothing about whether that motion characterizes coherent patterns such as biological or mechanical motion. In fact, our example in Figure 6A consists of visualized white noise. While disjoint distributions are passive in the sense that the observer need not be embedded in the flow of action, coherent movement itself is active and collective. Therefore, coherent movement is measured by considering the position of the observer during motion relative to other observers or objects in the environment.

In the example shown in Figure 6C, the random motions of balls in a chamber is compared to flocking (Pulliam et.al, 1973; Maruyama et.al, 2019) and swarming (Bonabeau et.al, 1999; Kennedy and Eberhart, 2001) behaviors in virtual agents. In fact, direct perceptual processing from an egocentric perspective by each agent in a group may actually enable the type of flocking exhibited in a boid model (Reynolds, 1987; Ward et.al, 2001). Each agent in the flock (observer) evaluates their position (heading) with respect to their previous position AND mean current heading of all observer neighbors. The greater correspondence between the agent's expectation and the group behavior (or minimized divergence), the higher the GI. Another example of coherent movement among individual agents is their observation of the environment around them. In the case of a shark, hydrodynamic phenomena (Zhou et.al, 2010) affect both sensation and the generation of movement through the environment. GI thus corresponds to laminar and turbulent flows, which in turn affect the sensation and movement against physical resistance (Wilga and Lauder, 2004).

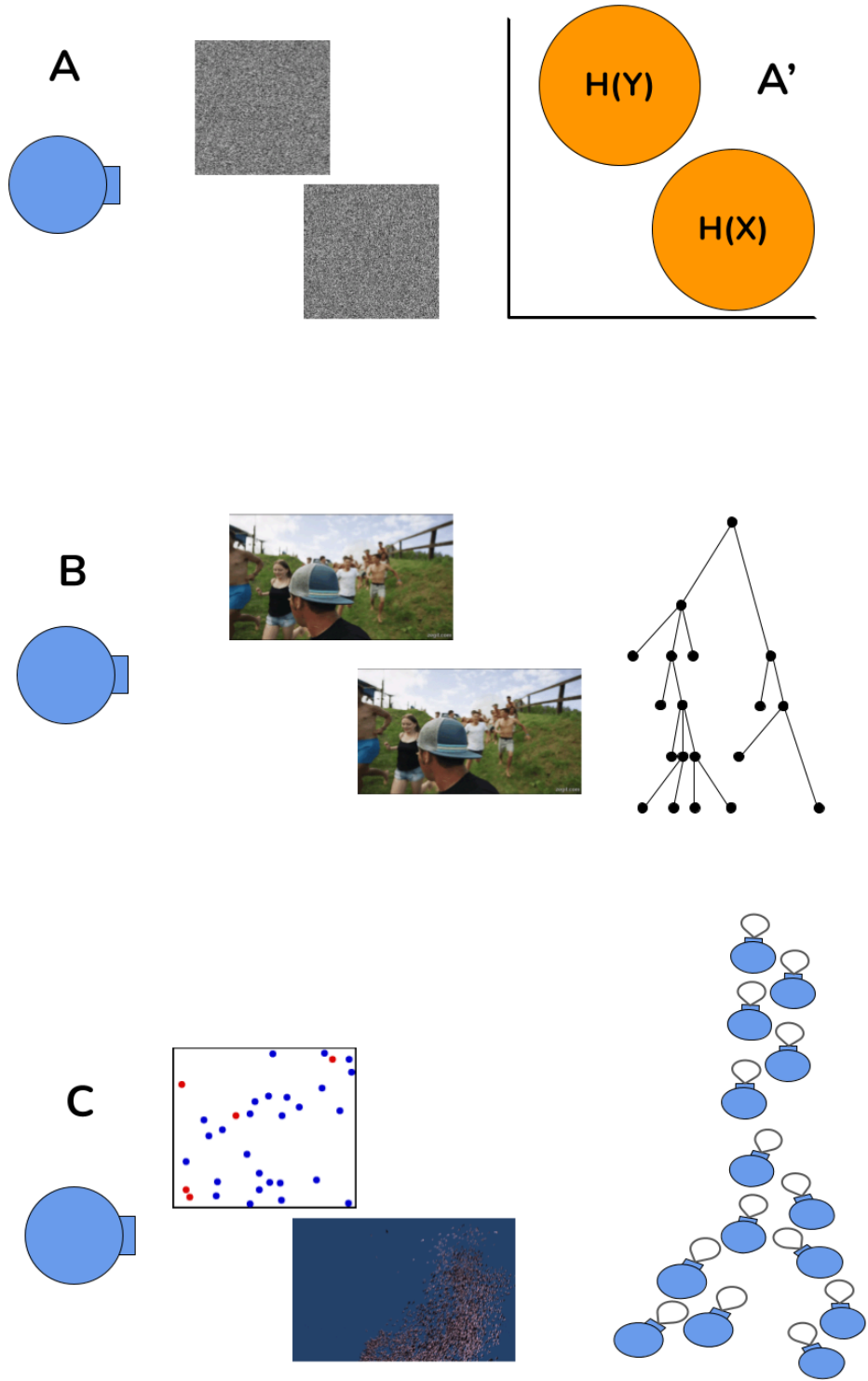

Figure 6. Three principles of GI, counterclockwise from top left: disjoint distribution (A), contingent action (B), and coherent movement (C). A: Disjoint GI is defined as the difference between H(X) and H(Y). This is comparable with Shannon Mutual Information (A'), which is defined as H(X,Y) = 0. B: an example of contingent action, where contingencies as represented by a directed tree structure. C: an example of coherent movement, where the previous heading is aligned with the current heading of the observer and the mean current heading observed among an agent's neighbors.

**Gratings and Bars**

We can also use stimuli as a means to demonstrate the fundamental capacities of nervous system information processing as captured by our three principles. For demonstration purposes, we will draw from the examples of gratings and bars. Colocating these two patterns against a neutral background provides affordances to the agent and increases the GI in its environment. In both of these cases, there is a high periodic structure unlike GI found in a natural scene. Their input distribution resembles a periodic binomial distribution similar to systems with high Shannon Information. Natural scenes also have no clear background, which is especially true in the case of coherent movement. To make the connection between periodic patterns and attentional processes, stripes are represented in the brain in the form of ocular dominance columns (Hubel and Wiesel, 1962). This relatively simple neural representation is consistent with the internal communication channel, and has connections to preattentive processes and attentional enhancement (Grossberg, 2001). In the developmental context, GI is restricted to certain developmental windows called critical periods that help to facilitate pattern formation (Katz and Crowley, 2002) and act as a resonant component of future information processing.

To make our toy model example more interesting, we can use gratings of different frequencies for both the background and foreground. This allows for GI to be evaluated as a set of differential distributions, where the background is subtracted from the foreground. In this case, environmental structure is represented by two competing distributions for which the differential strategy provides two possible evaluations of GI (for both foreground and background). Given a virtual reality stimulus, direct perceptual information viewed through the compound eye of the bee provides suppression of background and target cues enabling discrimination learning, while background and target cues interact to affect associative learning (Lafon et.al, 2021).

**What creates a nonlinear effect?**

In short: disturbances: either in the form of perceptual disruptions or discontinuities in the environment. Large changes in stimulus magnitude or the inability to perceive and integrate stimuli due to physiological shifts can contribute to perceptual disruptions, and is generally temporal in nature. Disturbances of sufficient magnitude or duration can lead to a generalized phase transition. GI can also serve to modulate and resolve logical contradictions. A view of information as sensory sampling across space and time might also enable internal mechanisms for multiple input/output (MIMO) processing, adaptive beamforming, and spatial smoothing (e.g. analogues from array signal processing) that disambiguate the directional orientation and source location of sensory signals (Zardi et.al, 2020 and Figures 7-9).

**Four-channel Embodiment**

Now that we have an appreciation for GI and its contrasts with Shannon Information, we can apply GI to a wider range of problems by using an example called four-channel embodiment. For a single sensory channel, we can observe various different architectural variations in response to these interactions with sensory information. These variations may prime these networks for later performance and learning benefits a phenomenon called developmental freedom (Alicea et.al, 2020). The addition of multiple sensory channels can add additional interactions to this basic framework, but with superadditive and suppressive effects. As both an acquisitional and morphological phenomenon, development is a key aspect of ecological approaches (Read and Szokolszky, 2018). In the case of dBVs, the developmental process provides insight into both how experience-dependent plasticity (Kleim and Jones, 2008) can enhance performance, and how different sensory cues in specific combinations can contribute to learning. These consequences are critical to understanding how GI can enhance or (in some cases) enable inference of a sensory channel.

**Multisensory Embodiment.** Given our interest in multisensory interactions, a four-channel analysis can serve to demonstrate how multisensory embodiment unfolds in an observer. In the following examples (Figure 7 and 8), a rabbit (Figure 7) and human performance in Augmented Reality (Figure 8) provides a means to embed multisensory GI into a single set of external and internal signals. These signal types represent the external and internal communication channels, respectively. The signals themselves have a specific orientation and strength as shown by the direction and length of the arrow. Figures 7 and 8 also provide a symbolic processing framework for multisensory interactions between the observer and their environment. For each sensory channel, a signal produces a directionality (from our spatial component of GI) and a perceived action (from our temporal component of GI). When all senses point in the same direction, they are said to be congruent. When different senses point in different directions, they are incongruent. This provides a symbolic processing framework that is consistent with the GI measurements shown in Figure 3.

From a broader perspective, there is an interesting relationship between disjoint distributions and multisensory embodiment of an agent that requires further study. In particular, the incongruence between stimuli shown in Figures 7 and 8 constitutes a disjoint distribution. While there is heterogeneity in the direction of action-perception across sensory modalities, this does not lead to breaks in perception. As consistency in the percept across modalities provides redundancies that lead to robustness, inconsistencies that are not too disruptive may serve as a mechanism for learned error correction. Indeed, our distinction between direct and indirect perception in Figure 3 is in line with this interpretation.

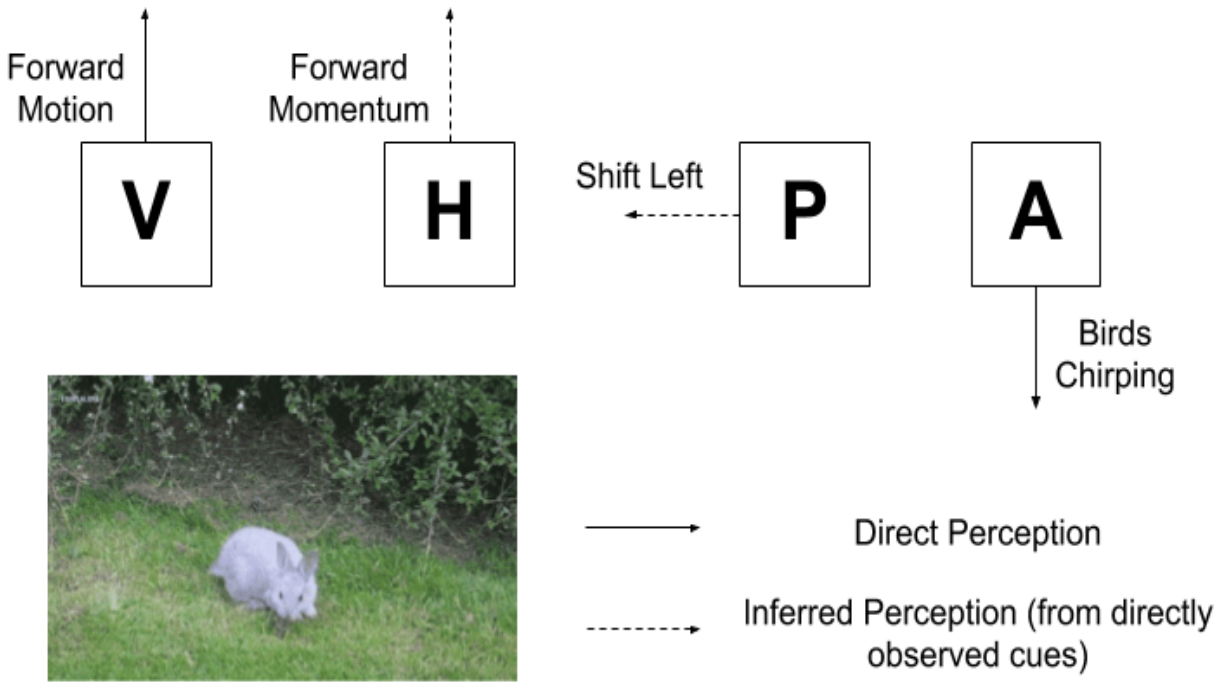

Figure 7. An example of four-channel embodiment: visual (V), haptic (touch - H), proprioception (P), and audition (A). The embodied observer is a rabbit hopping around a garden.

The roles of direct and indirect perception can also be demonstrated using an agent interacting with virtual reality. Virtual reality relies upon a computational simulation to create the illusion of a sensory environment. Once again, we can use our multisensory framework to understand what is going on in this interaction. Figure 8 demonstrates virtuality in the context of a human moving a virtual object with a wand and their upper body movements. In this case, a new form of direct perception called rotational momentum is introduced, present only in a single sensory modality.

**GI in a Multi-sensor Array.** The original mathematical description of GI assumed a single sensory apparatus, such as a single eye on an animal or a single haptic sensor on a robot. In most cases, GI is captured by an array of multiple sensors and then integrated during sensory transduction. An example of this is shown in Figure 9. In the case of an unweighted single sensor, we can consider the internal representation in terms of 1-dimension of time and 1-dimension of space. This will result in a blinkering of state (on and off) as sensory points are encountered. The movement that defines GI will be smeared in time as spatial information will be lacking. With multiple sensors, we can use different integration strategies to weight the different contributions of spatial diversity in the stimulus, similar to the center-surround weighting described previously.

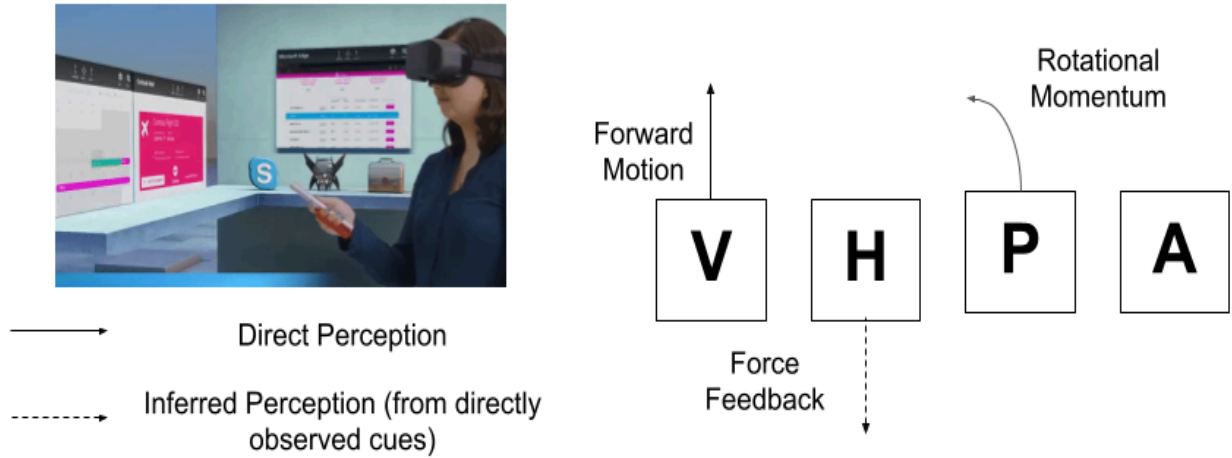

Figure 8. A secondary example of four-channel embodiment from human virtual environment navigation. Channels: visual (V), haptic (touch - H), proprioception (P), and audition (A).

**GI in Non-neuronal Systems**

Extending our model to information processing systems without a neuronal network, GI allows us to characterize both the internal and external communication channels of such a system. For example, in a Diatom cell, cells change their behavior with respect to light intensity (Cohn et.al, 2015). In bacteria, phototaxis and chemotaxis are the predominant mode of behavioral output (Ben-Jacob, 2008; Lan and Tu, 2016), and can be replicated using a small artificial neural network (Zhang et.al, 2021). These behaviors are driven by chemical and light gradients, respectively. GI involves both external and internal information processing mechanisms involving extracellular proteins and myosin fibers, respectively.

The motion of cells during migration and their arrangement in collectives are also an interesting application of GI. In this case, any single cell can exhibit collective motility during development. As part of a collective, this single cell can utilize coherent movement along with an integration of environmental cues as encoded input to estimate GI. GI is well suited to understanding the positional information cells use during development (Dubuis et.al, 2013; Tkacik and Gregor, 2021). In this case, the input is the distribution of gene expression values of a cell's biochemical milieu, and the model representing our internal communication channel is the mapping between the expected signal by location and the cell's autonomously-derived internal state.

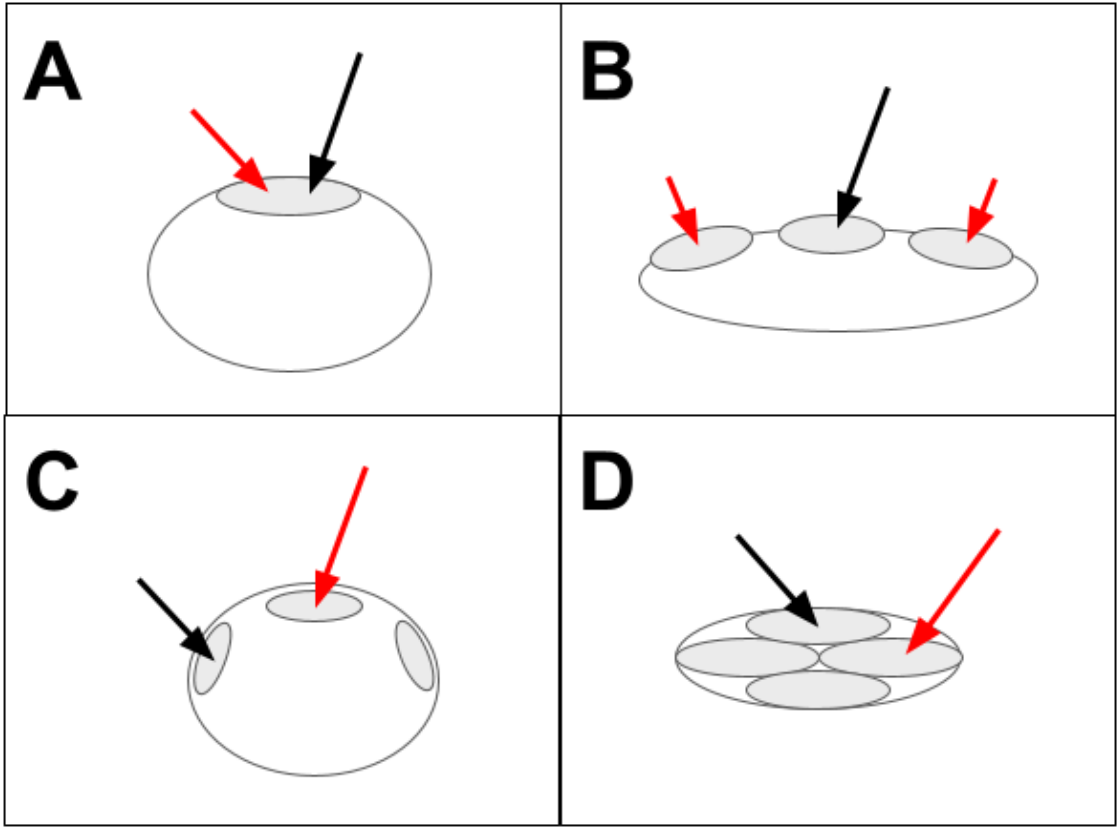

Figure 9. An agent that captures GI using a multi-sensor array and the resulting enrichment of spatial resolution. A: single sensor, B: parallel sensor array, C: semi-circular array, D: compound eye. Red arrow: input at edge of receptor field, Black arrow: input at center of receptor field.

## Theoretical Aspects of GI

### GI as Driver of Dynamical Collective Behavior

GI can also be used to mediate collective behavior between individual agents. The concept of coherent movement ($c$, Equ. 1) describes the perception of group movement by an individual. To achieve coordinated collective behavior, we can either measure the collective state of the group, or estimate the state of the group's interactions in both the participating agents and the coordinating stimulus itself. Coordination dynamics uses the order parameter ɸ to summarize the degree of synchrony in a phase space (Kelso, 1995; Tognoli, 2020). While this is a dynamical systems summary of the collective state, we might be interested in what exactly is coordinating our agents, who may produce a mean behavior but operate on heterogeneous information. Collective behaviors such as following a rhythm can be considered emergent properties that are produced through agentive interactions (Farrera and Ramos-Fernandez, 2021).

However those emergent properties do not emerge through magic: rather, they are the product of perception and action as a dynamical system. Sequences of perception can be thought of as a type of unsupervised program synthesis (Evans et.al, 2021), where sensory information contributes to the ability to predict and impute events by the internal model, and produces a response of appropriate magnitude. In GI, we propose that this difference is determined by the interaction of two parameters: $\tau$ (delay) and $g(i)$ (GI content). A phase space summarizing the approximate regions of these conditions are shown in Figure 10. In cases where both $\tau$ and $g(i)$ are low, we may observe a stagnation of cognitive state. This might prevent learning or result in non-responsiveness. For high $\tau$ and low $g(i)$, the cognitive state has a higher likelihood of collapse, where performance breaks down entirely. Medium values of both $\tau$ and $g(i)$ are essential for a sustained cognitive state resembling a flow state (Alameda et.al, 2022). While suboptimal environmental conditions can lead to drifting away from the sustainable region, this can also occur as a consequence of disease state or pathology (Ottinger et.al, 2021). Other sources of low-performance result from discontinuous shifts in perception (Shapiro et.al, 2010). During continuous visual tracking, objects moving towards the body show differences in resolution between different components of the sensory field. As a consequence of this and depending on the current cognitive state, performance can suffer in the ways demonstrated in Figure 10.

The regions of phase space are defined as collapse (across the range of $g(i)$ for moderate to high $\tau$), stagnation (low to moderate $g(i)$ combined with low $\tau$), sustainable (moderate to high $g(i)$ combined with moderate values of $\tau$), fast encoding (high values of $g(i)$ and low values of $\tau$), and high turnover defined as moderate to high $\tau$ and $g(i)$. High turnover results in a limited ability to encode memories, as much information is lost due to significant lag. On the other hand, fast encoding can occur when there is little lag but profuse information. Examples of fast encoding include fear conditioning, affordance-rich scenes, and highly-structured information characterized by $g(i)$ with a high $\lambda$ value).

**GI as a Complexity Metric.** GI can also provide novel views on how agents observe and respond to be embedded in complex systems. We will discuss three aspects of complex systems: instabilities and neural attunement. Instability (Mangalam, et.al, 2023) occurs when small variation in GI can lead to quick transitions and cascading failures in continuous behavior. The GI characterization of instabilities might result from large values of $\sigma$ (stochastic noise) or $\tau$ (temporal lag). As these conditions persist, the perceptual system can become unstable, leading to unpredictable failures of differing magnitudes. In defining an internal model, however, we can more closely examine and mitigate the specific conditions under which instabilities arise. Neural attunement is, as the name suggests, a means to reduce spontaneous failures. This would result in critical losses of dynamic GI during perceptual tracking or the acquisition of motion context. The latter involves controlling collisions using behavioral cues (Fajen and Devaney, 2006), and is characterized by smooth translations between our $c$ and $MV$ parameters. In a biological brain, this requires synchronization and alignment between neural populations, and can be controlled by oscillatory patterns (Shamay-Tsoory et.al, 2019; Levy et.al, 2021).

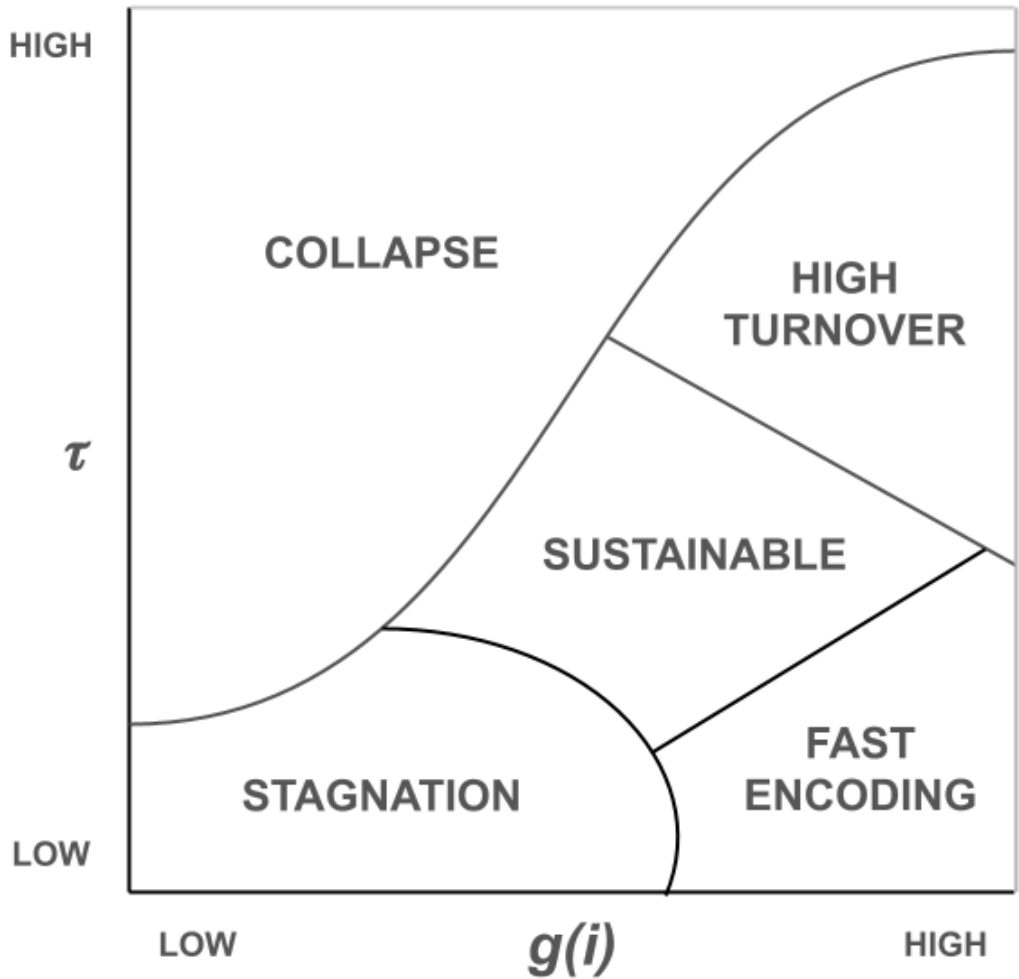

Figure 10. A conceptual phase space describing the different regimes of Gibsonian Information $g(i)$ versus delay $\tau$. Regions of the phase space are defined as: collapse, stagnation, sustainable, high turnover, and fast encoding.

**Local versus global information.** The measures and conceptualization of sensory information presented here is not assumed to occur at a single scale. In our example of collective behavior, phenomena such as decay, arrival, and sudden behavioral phase transitions provide a means for characterizing regulation over a dynamical trajectory. Yet many dynamical systems have processes that operate over multiple timescales. For example, bird flocks must both maintain local coherence while also steering around global obstacles using a strategy called marginal speed confinement (Cavagna et.al, 2022). The structure of GI theory, moving from a temporal extraction of environmental information to singular and collective proto-representations of that information, and finally to nervous system operations leading to movement and interactive behaviors, allows us to apply GI to both the context of the agent and the context of the wider world.

**Higher-order Complexity**

Another aspect of environmental information that is not explicit to our proposed measures is topology. Aside from temporal structure, the shape of Gibsonian information is an important higher-order feature of environmental information. This is particularly true for our measures of coherent movement ($c$, Equ. 1) and sensorium ($S$, Equ. 4). In its current form, $c$ is simplified to a scalar that can be applied as a series of planes, but could be explicitly $n$-dimensional for cases where GI is dominated by fractal or non-Euclidean geometries (Palatinus et.al, 2013; Albertazzi and Louie, 2016). The incorporation of touch and visual cues may exist in different dimensional regimes, touch being fractal (*3.x*-D) and vision being non-fractal (*3*-D). It may even be the case that visual perception is inherently a curved Riemennian space (Fernandez and Farell, 2009). In both cases, $c$ must be reformulated in terms of a specific spatial projection.

Having proposed measures and observational principles of Gibsonian Information, particularly in contrast to Shannon Information, we can finally consider the concept of channel capacity (Cover and Thomas, 1991) as it relates to GI. Channel capacity is a core concept in Shannon Information and psychophysics alike, and is similarly important in Gibsonian Information. In the realm of ecological perception (Gibson, 1979), implicitly defined capacities of sensory channels are used to enable an observer's ability to explore its environment (Lobo et.al, 2018). Channel capacity in the GI context is related to both cognitive capacity limits (Weems and Reggia, 2012) as well as transmission rate (Cover and Thomas, 1991). We can operationalize the dynamical channel capacity as either a continuous flow of bits or as a discrete ordering of bits, both represented as $g(i)$. Particularly in the continuous case, our mathematical formulation of flow rate and capacity ($f_{max}$) allows us to more completely represent $g(i)$ in the internal model. Flow rate can also be plugged into our representations of coherent movement ($c$) and the sensorium ($S$).

## Discussion

In this primer, we have provided a comprehensive discussion of the conceptual/theoretical terrain, mathematical models and measures, and naturalistic contexts of GI. We begin with a framing of ecological information in terms of its physical and biological bases, in addition to examples of how GI fits into ecological views of perception. While we provide three examples (prospective control, bouncing ball, and spatial integration), GI is meant as a general quantitative estimator of spatiotemporal structure in the perceptual environment. We then provide the mathematical definitions of key parameters (coherent movement, internal-related movement, flow rate capacity, decay, and the sensorium), applications of these parameters to the sensorimotor loop, and key comparisons between Gibsonian and Shannon Information. We then make connections to concepts in dynamical systems and dynamical cognition. Finally, the theoretical, naturalistic, and operational aspects of GI are presented with examples from multisensory perception (four-channel sensory embodiment in naturalistic contexts), computational agents (the Braitenberg Vehicles-based R, I, E model), and dynamical cognitive systems (higher-dimensional representations of the sensory world).

**Overarching Themes and Future Directions**

Occasionally, an agent can take multiple views of the world if the sensor array covers different parts of the same body (e.g. head and tail). This fundamental difference between different viewpoints allows for a demonstration of disjoint distributions and the measurement of GI. Particularly, a mental frame of reference shared either socially or within a central nervous system must integrate disparate forms of information. In the GI approach, differences between perspectives on the environment are generally summarized in a single parameter.

In a non-neuronal context, GI can be applied to pattern formation, particularly in biological systems such as developing embryos. As formulated, GI predicts that pattern formation is the inverse of pattern recognition. In Turing reaction-diffusion (Capone et.al, 2019; Van Gorder, 2021), chemical gradients can be reformulated as perceptual ones. For example, a perceptual gradient consisting of stripes or concentric circles against a background is quite similar to the formation of stripes and segments observed as resulting from pattern formation. Yet to account for this process without GI, this process seems to be vitalistic. As GI does not explicitly require neural representations, perception at the level of individual cells is simply an act of generating a temporal record of environmental experience with spatial context in the form of differential structure. Potentially high degrees of structure can be quantified within a reaction-diffusion gradient. Where boundaries are sharp, spatial acuity will tend to be narrow, and translates into high degrees of GI using our differential (spatially explicit) measure. Coincidentally, this is consistent with the positional information concept (Dubuis et.al 2013, Wolpert, 2016) in embryonic development.

Returning to the connection between perceptual information and thermodynamics, the characterization of GI in dynamical systems suggests that rather than a closed-loop feedback system, informational regulation more resembles a futile cycle. Futile cycles are native to metabolic physiology (Qian and Beard, 2006), and consist of closed loop feedback processes where the only product is energy dissipation. This type of organization allows for robustness to large fluctuations in molecule number (or pieces of information). GI also allows us to consider the role of externalization in cognitive agents. Externalizing cognition (Clark and Chalmers, 1998; Bocanegra et.al, 2019) can be understood through the encoding and decoding components of the schema shown in Figure 1. Externalization means that more of the representation of information occurs prior to the encoding stage. Therefore, the input distributions will tend to have more structure, and the increased number of affordances will likely take the form of complex exponential distributions characterized in Equation 1.

**Relationship to Alternate Approaches**

An alternative way to treat the probability distribution of environmental information is to map these points of information to a more general representation before encoding (see Figure 2). This allows us to build a mapping based on cybernetics and category theory (Cruttwell et.al, 2021). This spatially-explicit representation allows for the principle of coherent movement in space and contingent action in time. The former representation takes the form of a local-to-global map, and the latter is a tree-like structure. In terms of the resulting internal communication channel, we observe mappings between these representations that form a disjoint distribution. While the GI exists in these disjoint regions, there is also the potential for shared categories that also yield GI. Alternatively, there are connections between collective motion and non-classical logic (Krol et.al, 2021) that might serve to represent shared and disjoint categories in the internal communication channel. Such an approach might also allow us to extend the principle of disjoint categories in terms of category theory (Spivak, 2014).

Understanding direct perception as quantitative information may be quite effective for understanding phenomena like visual flow or in cases where the observers are slime molds and single cells. Yet in observers with complex nervous systems (such as bird flocks, hopping rabbits, or human roller coaster riders), direct perception may not be a sufficient means to explain how the sensory information translates into the inferential aspects of cognition and ultimately behavior (Fodor and Pylyshyn, 1981). Furthermore, direct perception is a non-representational view of the world, which limits its ability to identify and measure higher-level features of the environment (Hilario, 1997). This limits our ability to apply GI directly to behavior, at least in the form of quantitative measurements. But we may also be able to apply GI as the lower layer of a hybrid model of the observer in which GI is used to inform and/or constrain model components with greater representational capacity (e.g. Fodor and Pylyshyn, 1988; Feldman and Ballard, 1982).

# Appendix

Appendix A: list of quantitative parameters.

| Parameter | Common Usage | Definition |
|:---:|:---:|:---:|
| $f$ | $f_{max}$ | Flow rate and capacity |
| $MV$ | $MV_{MIN}, MV_{MAX}$ | Movement output |
| c | c | Coherent movement |
| $\vec{v}$ | $\vec{v}_v, \vec{v}_a, \vec{v}_t$ | Modal sensory velocity (vision, audition, touch) |
| $d$ | $d$ , $\frac{c-d}{MV}$ | Sensory decay |
| $VO$ | $PO_{vo}, \boldsymbol{\sigma}_{po}$ | Force production (volitional and noisy) |
| $S$ | $S$ , $S_t$ | Sensorium |
| $\tau$ | $\tau$ | Delay, temporal lag |